\documentclass[%
 reprint, 
superscriptaddress,
 amsmath,amssymb,
 aps, prd,
floatfix,
]{revtex4-2}

\usepackage{graphicx}
\usepackage{dcolumn}
\usepackage{bm}

\usepackage[utf8]{inputenc} 
\usepackage[T1]{fontenc}    
\usepackage{xcolor}
\definecolor{darkblue}{RGB}{46,48,147}
\usepackage[colorlinks=true,
            linkcolor=darkblue,
            urlcolor=darkblue,
            citecolor=darkblue]{hyperref}     
\usepackage{url}            
\usepackage{booktabs}       
\usepackage{amsfonts}       
\usepackage{nicefrac}       
\usepackage{microtype}      
\usepackage{lipsum}
\usepackage{xspace}
\usepackage{array}
\usepackage{adjustbox}
\usepackage[capitalise]{cleveref}
\usepackage{subfig}

\usepackage{array}
\usepackage{scalerel}
\usepackage{tikz}
\usetikzlibrary{svg.path}

\definecolor{orcidlogocol}{HTML}{A6CE39}
\tikzset{
  orcidlogo/.pic={
    \fill[orcidlogocol] svg{M256,128c0,70.7-57.3,128-128,128C57.3,256,0,198.7,0,128C0,57.3,57.3,0,128,0C198.7,0,256,57.3,256,128z};
    \fill[white] svg{M86.3,186.2H70.9V79.1h15.4v48.4V186.2z}
                 svg{M108.9,79.1h41.6c39.6,0,57,28.3,57,53.6c0,27.5-21.5,53.6-56.8,53.6h-41.8V79.1z M124.3,172.4h24.5c34.9,0,42.9-26.5,42.9-39.7c0-21.5-13.7-39.7-43.7-39.7h-23.7V172.4z}
                 svg{M88.7,56.8c0,5.5-4.5,10.1-10.1,10.1c-5.6,0-10.1-4.6-10.1-10.1c0-5.6,4.5-10.1,10.1-10.1C84.2,46.7,88.7,51.3,88.7,56.8z};
  }
}

\newcommand\orcidicon[1]{\href{https://orcid.org/#1}{\mbox{\scalerel*{
\begin{tikzpicture}[yscale=-1,transform shape]
\pic{orcidlogo};
\end{tikzpicture}
}{0}}}}

\newcommand{\ptvecmiss}{\ensuremath{{\vec p}_{\mathrm{T}}^{\kern1pt\text{miss}}}\xspace}

\newcommand{\GEANTfour}{{\textsc{Geant4}}\xspace}

\newcommand{\pythia} {{\textsc{Pythia8}}\xspace}

\newcommand{\pandora} {{\textsc{PandoraPfa}}\xspace}
\newcommand{\hitpf} {{\textsc{HitPf}}\xspace}

\newcommand{\akfourchs}[1]{AK4-CHS\xspace}
\newcommand{\akfourpuppi}[1]{AK4-PUPPI\xspace}

\makeatletter
\long\def\@makecaption#1#2{%
  \vskip\abovecaptionskip
  \sbox\@tempboxa{#1: #2}%
  \ifdim \wd\@tempboxa >\hsize
    #1: #2\par
  \else
    \global \@minipagefalse
    \hb@xt@\hsize{\box\@tempboxa\hfil}%
  \fi
  \vskip\belowcaptionskip}
\makeatother

\begin{document}

\preprint{APS/123-QED}

\title{\textbf{End-to-end event reconstruction for precision physics at future colliders} 
}%


\author{Dolores Garcia\orcidicon{0000-0002-0120-8757}}
 \email{Contact author: dolores.garcia@cern.ch}
\affiliation{%
    \href{https://ror.org/01ggx4157}{European Center for Nuclear Research (CERN)}, Geneva 1211, Switzerland 
}%
\author{Lena Herrmann}
\affiliation{%
    \href{https://ror.org/01ggx4157}{European Center for Nuclear Research (CERN)}, Geneva 1211, Switzerland 
}%

\author{Gregor Krzmanc}
\affiliation{%
    \href{https://www.stanford.edu/}{Stanford University}, Stanford, CA 94305, USA 
}%

\author{Michele Selvaggi}
\affiliation{%
    \href{https://ror.org/01ggx4157}{European Center for Nuclear Research (CERN)}, Geneva 1211, Switzerland 
}%

\begin{abstract}
Future collider experiments require unprecedented precision in measurements of Higgs, electroweak, and flavour observables, placing stringent demands on event reconstruction. The achievable precision on Higgs couplings scales directly with the resolution on visible final state particles and their invariant masses. Current particle flow algorithms rely on detector specific clustering, limiting flexibility during detector design. Here we present an end-to-end global event reconstruction approach that maps charged particle tracks and calorimeter and muon hits directly to particle level objects. The method combines geometric algebra transformer networks with object condensation based clustering, followed by dedicated networks for particle identification and energy regression. Our approach is benchmarked on fully simulated electron positron collisions at FCC-ee using the CLD detector concept. It outperforms the state-of-the-art rule-based algorithm by 10--20\% in relative reconstruction efficiency, achieves up to two orders of magnitude reduction in fake-particle rates for charged hadrons, and improves visible energy and invariant mass resolution by 22\%. By decoupling reconstruction performance from detector-specific tuning, this framework enables rapid iteration during the detector design phase of future collider experiments.

\end{abstract}

\maketitle

\clearpage

\section{Introduction}
\label{sec:introduction}

Reconstructing individual particles from the raw signals of a particle detector is the foundational step connecting experimental measurements to physical observables, and the accuracy of this reconstruction directly drives the sensitivity of fundamental physics measurements. The Future Circular Collider (FCC) has been proposed~\cite{FCC:2025lpp, deBlas:2025gyz} as a next-generation facility to follow the Large Hadron Collider (LHC) at CERN. Its first stage, FCC-ee, will collide electron-positron beams at the $Z$, $WW$, $ZH$, and $t\bar{t}$ thresholds with large instantaneous luminosity. Experiments at the FCC-ee aim to measure Higgs couplings, electroweak parameters, and flavour observables with exceptional precision. The achievable sensitivity on key Standard Model parameters is directly driven by the resolution with which visible final state particles and their invariant masses can be reconstructed. Rare hadronic Higgs decays such as $H\rightarrow c\bar{c}$ and $H\rightarrow s\bar{s}$ are particularly challenging, since their relative uncertainty scales are strongly sensitive to the visible mass resolution $\sigma_m$. More generally, hadronic final states at FCC-ee feature large particle multiplicity and geometrical overlap, placing stringent demands on detector design and reconstruction algorithms.

Particle flow (PF) reconstruction addresses these challenges by optimally combining tracking and calorimeter and muon information to reconstruct individual final-state particles with the best achievable resolution. First developed at LEP~\cite{ALEPH:1994ayc, DELPHI:1995dsm} and extended to $ep$~\cite{Wing:2000tt} and hadron colliders~\cite{CMS:2009nxa, CMS:2017yfk, ATLAS:2017ghe}, PF will be the baseline approach for future collider projects, including FCC-ee~\cite{Bacchetta:2019fmz}, CEPC~\cite{Ruan:2013rkk}, ILC~\cite{Behnke:2013lya, Breidenbach:2021sdo}, and CLIC~\cite{Linssen:2012hp}. Classical implementations share a common structure: iterative calorimeter clustering followed by track cluster association and neutral particle identification~\cite{buskulic1995performance, gray2016challenges, thomson2009particle}. The \pandora algorithm~\cite{thomson2009particle}, currently the baseline for FCC-ee, achieves excellent jet energy and visible mass resolution but relies on extensive detector-specific tuning, limiting flexibility during the detector design and optimisation phase.

Recent works have explored machine learning (ML) approaches to PF reconstruction~\cite{DiBello:2020bas, Kieseler:2020wcq, Pata:2021oez, Mokhtar:2023fzl, qasim2021multi, DiBello:2022iwf, Kakati:2024dun}. The MLPF algorithm~\cite{Pata:2021oez} uses graph neural networks to regress particle properties from pre-clustered inputs and improves jet resolution at linear collider detectors and CMS~\cite{Pata:2022wam, Mokhtar:2023fzl}. Other approaches reformulate PF as hypergraph learning~\cite{DiBello:2022iwf, Kakati:2024dun} or employ object condensation for calorimeter clustering~\cite{Kieseler:2020wcq, qasim2021multi}. However, these methods either rely on detector specific clustering or address only partial aspects of global event reconstruction, limiting their applicability to novel detector concepts and design optimization.

\begin{figure*}[htbp]
    \centering
    \includegraphics[width=0.9\textwidth]{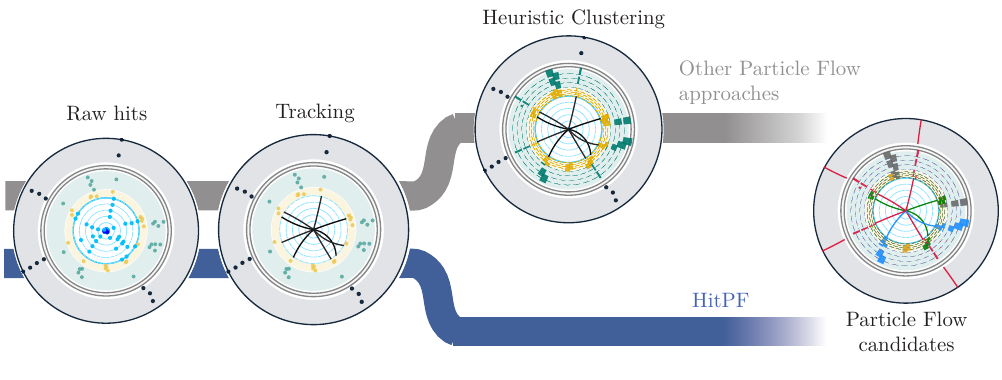}
    \caption{Overview of the \hitpf reconstruction pipeline compared to conventional Particle Flow algorithms. Traditional approaches rely on sequential, hand-tuned stages that first cluster calorimeter hits before associating them with tracks and assigning particle properties. \hitpf bypasses these intermediate steps, reconstructing final state particles directly from detector hits in a single end-to-end trainable model.}
    \label{fig:hitpfworkflow}
\end{figure*}

Rather than relying on such intermediate reconstruction steps, we aim to learn a global event representation directly from low-level detector information. This approach introduces fundamental challenges as individual particles must be uniquely identified within high multiplicity, highly collimated electromagnetic and hadronic showers. Concretely, a set of $N$ detector hits
must be associated to $M$ particles, where $N \gg M$. This approach differs
fundamentally from the above approaches~\cite{DiBello:2020bas,
Kieseler:2020wcq, Pata:2021oez, Mokhtar:2023fzl, qasim2021multi,
DiBello:2022iwf, Kakati:2024dun}, where the inputs are pre-formed clusters
with $N \sim M$. In such an environment, neutral particle reconstruction is particularly critical, as insufficient clustering or inaccuracies in track-cluster geometrical matching can generate fake energy that degrades energy and mass resolution. In contrast, merging a neutral hadron with a nearby charged hadron shower prevents exploitation of the superior track momentum measurement, and instead forces reliance on the inferior calorimetric energy estimate. Additionally, any such approach must achieve computational scalability despite large hit multiplicities, and ultimately match or exceed the performance of extensively validated rule based algorithms.

Here we present a unified solution, called \hitpf, that maps calorimeter hits and charged particle tracks directly to physics level particle objects, avoiding intermediate clustering stages and detector specific heuristics (Figure~\ref{fig:hitpfworkflow}), thereby enabling rapid development cycles for new detector concepts and geometries. The method reconstructs particles in two stages. First, a geometric algebra transformer network combined with object condensation clustering assigns hits to particle candidates. Second, dedicated regression networks determine the properties of each candidate: energy and particle type, classified into one of five categories (charged hadron, photon, neutral hadron, electron, muon). The optimization procedure enhances resolution and reconstruction efficiency while minimizing particle misidentification. Our algorithm is benchmarked against the \pandora algorithm using fully simulated $e^+e^-$ hadronic events at $\sqrt{s} = 91$~GeV in the FCC-ee CLD detector~\cite{bacchetta2019cld}.

\section{Methods}
\label{sec:methodology}

The \hitpf algorithm reconstructs particles from low level detector inputs. We frame particle reconstruction from hits as an instance segmentation problem \cite{kolodiazhnyi2024oneformer3d, sun2023superpoint}, where for a given point cloud (set of hits), the task is to identify objects (particles) and assign a unique identifier and properties to each one. Conceptually, this differs largely from previous ML-based particle flow approaches such as Ref.~\cite{pata2021mlpf}, where the inputs are clusters and tracks and one output per input element is predicted.   

The \hitpf algorithm is composed of two stages.
In the first stage, a geometric algebra transformer network  encodes calorimeter hits and tracks into a learned latent representation, a low-dimensional space that captures the essential features of each input, where hits belonging to the same particle cluster together. A density peak clustering algorithm extracts particle candidates from this latent space. In a second stage, dedicated networks perform particle identification, classifying candidates into one of five categories (charged hadron, photon, neutral hadron, electron, muon), and reconstruct the particle energy. The final output is a list of reconstructed particles with calibrated four momenta and particle type assignments. 
We describe the dataset, training targets, loss function, and network architecture below.

\begin{figure*}[htbp]
    \centering
    \includegraphics[width=\textwidth]{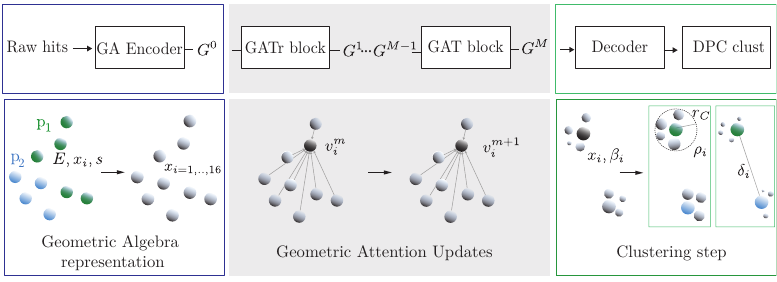}
    \caption{Architecture of the candidate determination step. The network takes as input detector hits from all sub-detectors and outputs particle candidate clusters. Left (blue): the encoder maps hit features into multivectors in a geometric algebra representation. Centre (grey): a Geometric Transformer performs $M$ GATr message-passing rounds over latent graphs $G_0, \ldots, G_M$; shaded blocks indicate trainable components. Right (green): the decoder extracts point and scalar components from the output multivectors and performs density-peak clustering (DPC). Each hit is assigned a local density $\rho$ and a distance $\delta$ to the nearest node of higher density; cluster centres are identified as nodes with simultaneously high $\rho$ and high $\delta$. 
     }
    \label{fig:model_structure}
\end{figure*}   

\subsection{Dataset and simulation}
\label{sec:dataset}

We produce a dataset of 1.1M $e^+e^-$ collision events ($Z\rightarrow
q\bar{q}$, $q=u,d,s$) at the $Z$ pole centre-of-mass energy
$\sqrt{s}=91$~GeV, hereafter referred to as the \textit{physics dataset}.
The final state consists of two back-to-back jets with high particle
multiplicity, providing a challenging benchmark for particle flow
reconstruction: the dense hadronic environment produces frequent shower
overlaps and a broad spectrum of neutral and charged particles across a wide
energy range. To complement the physics dataset, where isolated leptons and
prompt photons are underrepresented, we produce a second dataset of 200k
events, referred to as the \textit{gun dataset}, in which 10 single
particles per event are generated in isolation, drawn uniformly from the five
particle categories with energies in $[0.5, 50]$~GeV.
Events are generated with \pythia~\cite{Bierlich:2022pfr} and propagated through the CLD detector geometry (\texttt{CLD\_o2\_v05})~\cite{CLDConfig2023} using \GEANTfour~(v11.0.2)~\cite{GEANT42003,GEANT42006,GEANT42016}. Reconstruction uses the \textsc{Key4HEP} software stack~\cite{ganis2022key4hep} with the \textsc{Marlin} framework~\cite{Gaede:2006pj}. The \pandora algorithm~\cite{Marshall:2012hh,Marshall:2012ry,Marshall:2015rfa}, implemented within the CLD reconstruction pipeline~\cite{bacchetta2019cld}, provides the baseline for comparison.

The inputs consist of reconstructed tracks and calorimeter hits from the ECAL, HCAL, and muon system. Each input element is characterised by $(E, \vec{x}, s)$, with $s \in \{\textrm{Track}, \textrm{ECAL}, \textrm{HCAL}, \textrm{Muon}\}$ denoting the subdetector. For tracks, the position corresponds to the track state at the calorimeter entrance and $E$ is the track energy measured at the vertex assuming a charged pion mass. 
Hits from all subdetectors share the same input format. Events contain, on
average, 25 particles, 600 hits per particle, and 10 tracks. The physics
dataset is split into 900k events for training, 100k for validation, and
100k for testing, and is used for all results presented in
Section~\ref{sec:results}. The single-particle dataset is used to train the
property regression networks, as described in
Section~\ref{sec:property_regression}. The raw dataset in EDM4HEP format and
the generation code are publicly available at~\cite{Garcia_2026_18751314},
and the ML-ready datasets at~\cite{Garcia_2026_18749298}.

\subsection{Training targets}
\label{sec:loss}


The reconstruction targets are defined as Monte Carlo particles at the entrance of the calorimeter. This choice provides the network with unambiguous training labels corresponding to distinct detector signatures. For example, an unconverted photon produces a single electromagnetic cluster, while a converted photon produces an electron-positron pair with associated tracks and clusters. Using stable generator level particles as targets, as done in other approaches~\cite{mokhtar2025fine} would require the network to learn that both signatures correspond to the same truth label, leading to ambiguous labels for supervision. As a result of the chosen target, subsequent reconstruction steps are required to obtain analysis level objects, such as electron bremsstrahlung recovery or photon conversion reconstruction, but this factorization simplifies the learning task and improves reconstruction performance. The baseline reconstruction algorithm, \pandora, also solves the low-level task of finding resolvable clusters in the detector, and therefore also requires these subsequent algorithms. This factorised, multi-stage approach is standard practice in both classical and ML-based particle flow reconstruction.

\subsection{Architecture and inference workflow}
\label{sec:architecture}

We now describe the two stages of the \hitpf pipeline in detail: hit clustering (Section~\ref{sec:gatr}) and property regression (Section~\ref{sec:property_regression}). A schematic overview of the clustering stage is shown in Figure~\ref{fig:model_structure}.

\subsubsection{Clustering of hits }
\paragraph{Geometric algebra transformer}
\label{sec:gatr}

\begin{figure*}[htbp]
    \centering
    \includegraphics[width=0.99\textwidth]{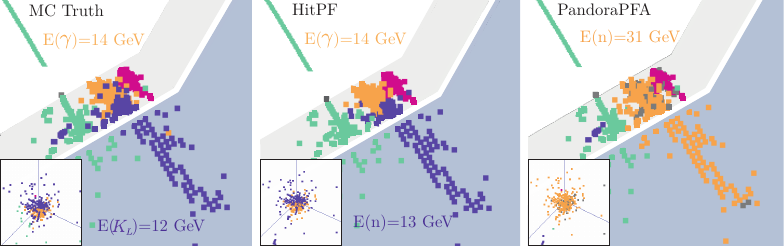}
    
    \caption{Section of a $Z\rightarrow q\bar{q}$ event showing the target hit
    assignment (left), the \hitpf reconstruction (center), and the \pandora
    reconstruction (right). The event showcases two overlapping showers
    originating from a $K_L$ and a photon. The colour assigned to each hit for
    \hitpf and \pandora corresponds to the MC particle contributing the
    largest hit count to the reconstructed cluster. The bottom subplot in each
    panel shows the two showers projected along the shower axis.}
    \label{fig:neutrals_figure}
\end{figure*}
The clustering backbone uses the geometric algebra transformer architecture (GATr)~\cite{brehmer2023geometric}, which embeds detector hits as geometric objects in the projective geometric algebra $\mathcal{G}_{3,0,1}$. We embed the hit coordinates in the point component of the multivector, $E$ in the scalar component and use the additional scalar inputs to include the subdetector discrete identifier as a binary-encoded vector in which only the entry corresponding to the given subdetector is set to 1 and all others are 0.
 This representation also accommodates planes, vectors, or 3D objects, which may prove useful for other detector technologies, but are not used in this work. Operations between multivectors allow geometric manipulation of the inputs, including projections, angles between vectors (inner product), subspaces spanned by vectors (wedge product), and magnitudes such as lengths, areas, or volumes. Typical GNN or transformer architectures use vectors of scalar features, requiring the network to implicitly learn geometric relations without direct access to this information. Instead, this representation provides a strong inductive bias for spatial detector data while maintaining computational efficiency through dot-product attention. 


The network consists of 10 GATr blocks, each composed of three layers: linear layers with equivariance constraints,  attention layers with $E(3)$ equivariance, and geometric bilinear layers that construct new geometric types. We set 16 hidden channels for the multivector layers and 64 for the scalar layers, with 8 heads. This configuration has a total of 924,000 trainable parameters.

The network is trained using the object condensation loss~\cite{Kieseler:2020wcq}, which enables the reconstruction of a variable number of particles by training the network to cluster hits belonging to the same particle in the output latent space while separating hits from different particles. Below we provide an intuitive description of the loss and refer the reader to~\cite{Kieseler:2020wcq} for a detailed formulation.

The network outputs coordinates $x_i$ (extracted from the point component) and condensation scores $\beta_i$ (from the scalar component) for each hit $i$, these constitute a learned latent representation. Let us refer to the outputs of the network for each hit as a node. Ideally, each particle has one node with a high condensation score, while the remaining nodes belonging to that particle have low scores and cluster around it in the latent space. The object condensation loss achieves this effect by defining an effective charge derived from the condensation score, and introducing an attractive potential toward the highest-charge node within each object (the condensation point). Each node is therefore attracted to its own condensation point and repelled by the condensation points of other particles.

\paragraph{Clustering}
\label{sec:clustering}
At inference time, particle candidates are extracted from the learned latent space using a variant of the density peak clustering (DPC) algorithm~\cite{rodriguez2014clustering}. DPC identifies cluster centres as points with high local density that are distant from any point with higher density. The local density is computed as
\begin{equation}
    \rho_i = \sum_j E_j \, \chi(d_{ij} - d_c),
\end{equation}
where $E_j$ is the energy of each point, $d_c$ is a cutoff distance, set to $0.1$, and $\chi(x) = 1$ if $x < 0$ and zero otherwise. 
The quantity $\delta_i$ measures the minimum distance to any point with higher density:
\begin{equation}
    \delta_i = \min_{j:\, \rho_j > \rho_i} (d_{ij}).
\end{equation}
Cluster centers are identified as points with both high $\rho$ and high
$\delta$, with thresholds set at $\rho_{\min} = 0.05$~GeV, corresponding to
the minimum particle energy of interest, and $\delta_{\min} = 0.5$. All remaining points within distance $d_p = 0.5$ are assigned to the nearest center.

We choose DPC over alternatives such as condensation-score-based
selection~\cite{Kieseler:2020wcq} or
HDBSCAN~\cite{mcinnes2017hdbscan} for its natural suppression of fake
particles. Fake clusters typically arise from two failure modes: splitting a hadronic
shower produces fragments with high $\rho$ but low $\delta$, while isolated
low energy deposits exhibit low $\rho$ but high $\delta$. By requiring
cluster centers to satisfy both criteria simultaneously, this algorithm is effective against both
sources.

Clusters produced by DPC consist of sets of calorimeter, muon hits and track nodes. Ideally, clusters produced by charged particles would contain one track node and a set of hit nodes, and clusters produced by neutral candidates would only contain hit nodes. In rare cases, however, a cluster can be associated with the wrong track; we therefore impose a consistency check to remove spurious track assignments. The track is removed from the cluster if the difference between the track and the cluster is four standard deviations away from the resolution of the worse resolution in the sub-detector,   $\frac{E_{\textrm{cluster}}-p_{\textrm{track}}}{p_{\textrm{track}}}> 4\frac{a}{\sqrt{E_{\textrm{cluster}}}}$. In this case, we take $a$ as the stochastic term of the HCAL calorimeter. This procedure affects approximately 2\% of all tracks. If multiple tracks remain in a cluster, the
one with the smallest energy difference is kept.  Unlike traditional particle flow algorithms,
which perform iterative cluster splitting in such cases, \hitpf relies on
the high clustering efficiency of the algorithm and assumes each cluster
corresponds to a single particle.

Figure~\ref{fig:neutrals_figure} illustrates the reconstruction of a
representative region of a $Z \rightarrow q\bar{q}$ event containing two
overlapping showers from a $K_L$ and a photon. The target hit assignment
(left) shows two distinct particles whose calorimeter deposits
partially overlap. \hitpf (centre) correctly resolves the two showers into
separate clusters, while \pandora (right) merges them into a single
neutral hadron with a reconstructed energy of 31~GeV, significantly
exceeding the sum of the true energies (12 and 14~GeV). This excess arises
not only from the merging itself, but also from the application of a single
hadronic energy calibration to a mixed electromagnetic-hadronic cluster: since
the calorimeter response differs for electromagnetic and hadronic showers,
the incorrect compensation further biases the energy estimate. This example
illustrates how shower merging in dense environments degrades both particle
reconstruction efficiency and energy measurement. The quantitative impact on reconstruction
efficiency, fake rates, and energy resolution is assessed in
Section~\ref{sec:results}.

\subsubsection{Property regression}
\label{sec:property_regression}

Particle identification and energy regression are performed by separate networks, each operating on collections of hits and tracks assigned to the individual particle candidates in the clustering step. For each candidate cluster, the nodes (hits and potentially a track) are first passed into a GATr network with 3 blocks, 2 hidden multivector channels, and 4 hidden scalar channels. The 16-dimensional multivector outputs from GATr are then summed up into a single embedding vector for each reconstructed cluster. 

In addition to the GATr embeddings, a set of 11 aggregate features is computed per candidate cluster: the fraction of energy in ECAL hits, the fraction of energy in HCAL hits, the total number of hits, the number of tracks, the track momentum, the track fit $\chi^2$, the variance of ECAL hit energies, the variance of HCAL hit energies, the total hit energy sum, the number of muon system hits, and the muon system hit energy sum.

The 11 aggregate features are concatenated to the 16-dimensional geometric representation of the candidate clusters, and finally passed through a 3-layer neural network with an embedding dimension of 64 and ReLU activation function. The energy regression and particle identification networks have the same architecture, except for the last layer. The energy correction returns a single scalar representing the energy of the particle. We use separate weights for particle identification networks applied to particle candidates containing a track node in the cluster (charged candidates), and for the neutral particle candidates. Therefore, the network outputs class probabilities of the possible particle types (charged: charged hadron, $e^{\pm}$, $\mu^{\pm}$; neutral: photon, neutral hadron). The reconstructed particle ID is defined as the class with the highest predicted probability.

The particle identification network is trained with cross-entropy loss against the target particle type. 
The energy regression network is trained with L1 loss against the true Monte Carlo particle energy. 

For charged particles, energy and direction are obtained from the associated track momentum at the interaction vertex. Additionally, charged candidates with momentum $p < 1$~GeV
identified as muons are reassigned to charged hadrons, since at these
momenta the tracking information alone cannot distinguish the two species and
the network prediction is effectively unconstrained. For neutral particles, the direction is computed from the energy-weighted average hit positions from the interaction vertex. 

\subsubsection{Training}
\label{sec:training}

The candidate determination (clustering) and property regression stages are trained separately. The candidate determination is trained using the $Z\rightarrow q\bar{q}$ dataset, as it features more complex shower geometries and higher particle multiplicities. The property regression is trained on the second dataset, generated using a particle gun, as it provides a more balanced representation of particle types  which are underrepresented in the first dataset as discussed in Section. \ref{sec:dataset}. The clustering model is trained for 10 epochs on 4 NVIDIA H100 GPUs with batch size 20, using the Adam optimiser~\cite{adamw} with cosine annealing learning rate scheduling from $10^{-3}$ to $10^{-6}$. The model is trained for 48 hours.

For energy regression, the epoch minimizing the validation L1 loss is selected. The property regression networks are trained for 5 epochs with the same optimizer configuration, selecting the epoch with the highest validation accuracy for particle identification. The training and inference pipeline is implemented in PyTorch \cite{pytorch2019}.

\begin{figure*}[htbp!]
    \includegraphics[width=0.99\textwidth]{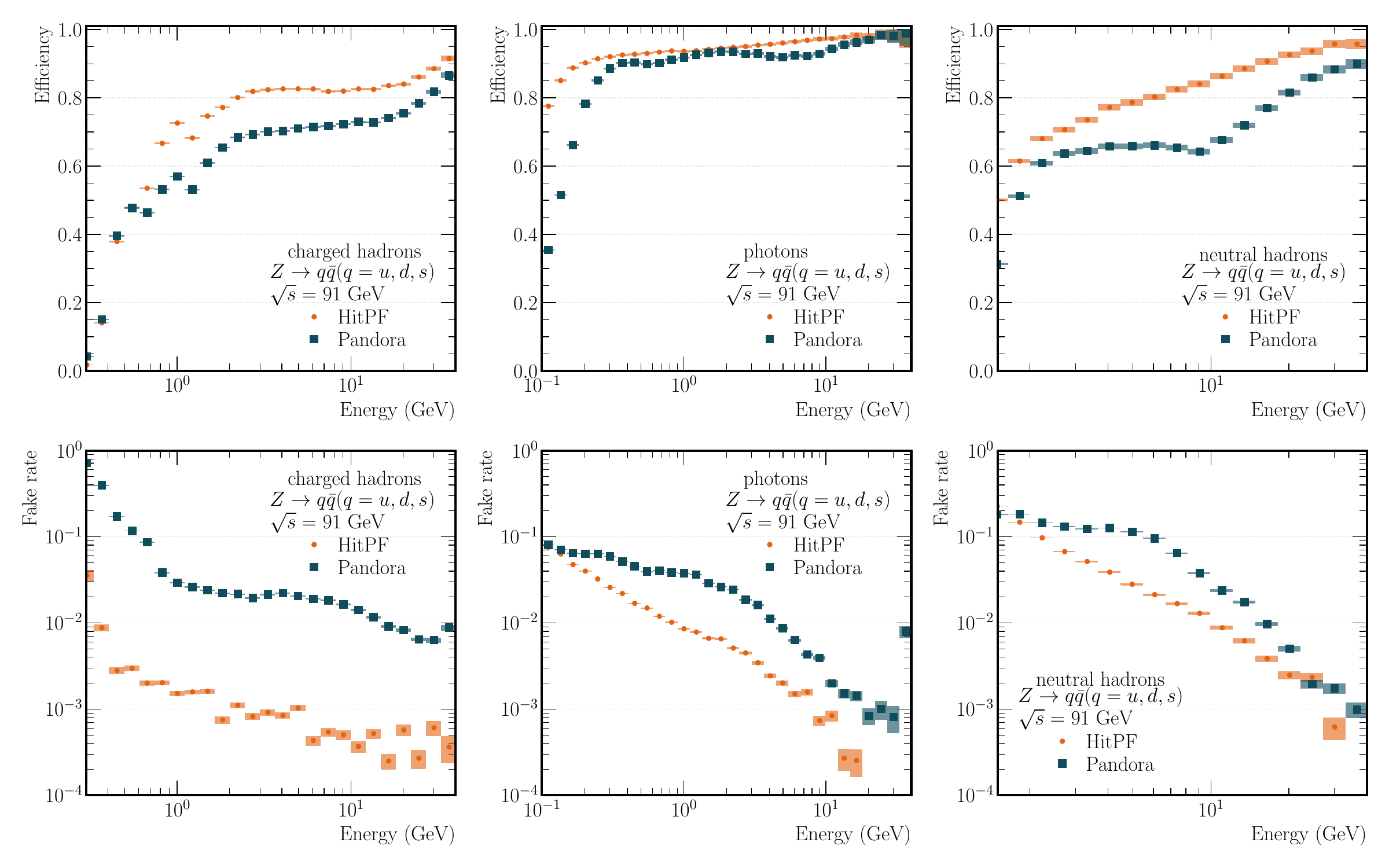}
    \caption{Particle identification performance as a function of energy for charged hadrons (left), photons (centre), and neutral hadrons (right). Top row: reconstruction efficiency, defined as the fraction of target particles reconstructed with the correct label, shown as a function of the matched MC particle energy. Bottom row: fake rate, defined as the fraction of reconstructed candidates without a matched target particle, shown as a function of the energy of the reconstructed particle.}
    \label{fig:efficiency_fake_rate}
\end{figure*}
We evaluate \hitpf on 100,000 test events from the physics dataset and compare against \pandora using particle-level and event-level metrics.

\section{Results}
\label{sec:results}

\subsection{Particle Identification}

The goal of particle identification (PID) is to correctly classify each reconstructed candidate into one of five categories: charged hadrons, photons, neutral hadrons, electrons, and muons, while maximizing reconstruction efficiency and minimizing the misidentification probability.

We define the PID efficiency as the fraction of target particles successfully reconstructed with the correct label, and the fake rate as the fraction of reconstructed clusters of a given predicted PID class without a matched target particle.

Figure~\ref{fig:efficiency_fake_rate} shows the efficiency and fake rate as a function of particle energy for charged hadrons, photons, and neutral hadrons. Electrons and muons are discussed via the confusion matrices (see discussion below), where all five classes are presented. 

For charged hadrons, \hitpf achieves up to 20\% higher efficiency in the 1-10~GeV range compared to \pandora, with a fake rate reduced by up to two orders of magnitude across all energies. The efficiency is limited by the upstream track reconstruction, which is external to \hitpf. However, a large fraction of the lost tracks are still reconstructed as clusters, as \hitpf recovers them with higher clustering efficiency than \pandora, as shown in Figure~\ref{fig:clustering_metrics_appendix} in Appendix~\ref{app:clustering_metrics}.

For photons, \hitpf provides approximately 5\% higher efficiency at high energy, with improvements up to a factor of two for $E < 1$~GeV. The fake rate is reduced by up to an order of magnitude in the 1--10~GeV range.

For neutral hadrons, \hitpf reaches up to 20\% higher efficiency around $E = 10$~GeV, together with an almost factor 5 reduction in fake rate. In this region, \hitpf reconstructs complex shower topologies without splitting clusters into sub-components, thereby reducing fake clusters. The fraction of total energy assigned to fake candidates is shown Figure~\ref{fig:fake_energy} in Appendix~\ref{app:clustering_metrics}, confirming that the overall picture remains consistent with the trends discussed above.

\begin{figure*}[ht]
    \includegraphics[width=0.8\textwidth]{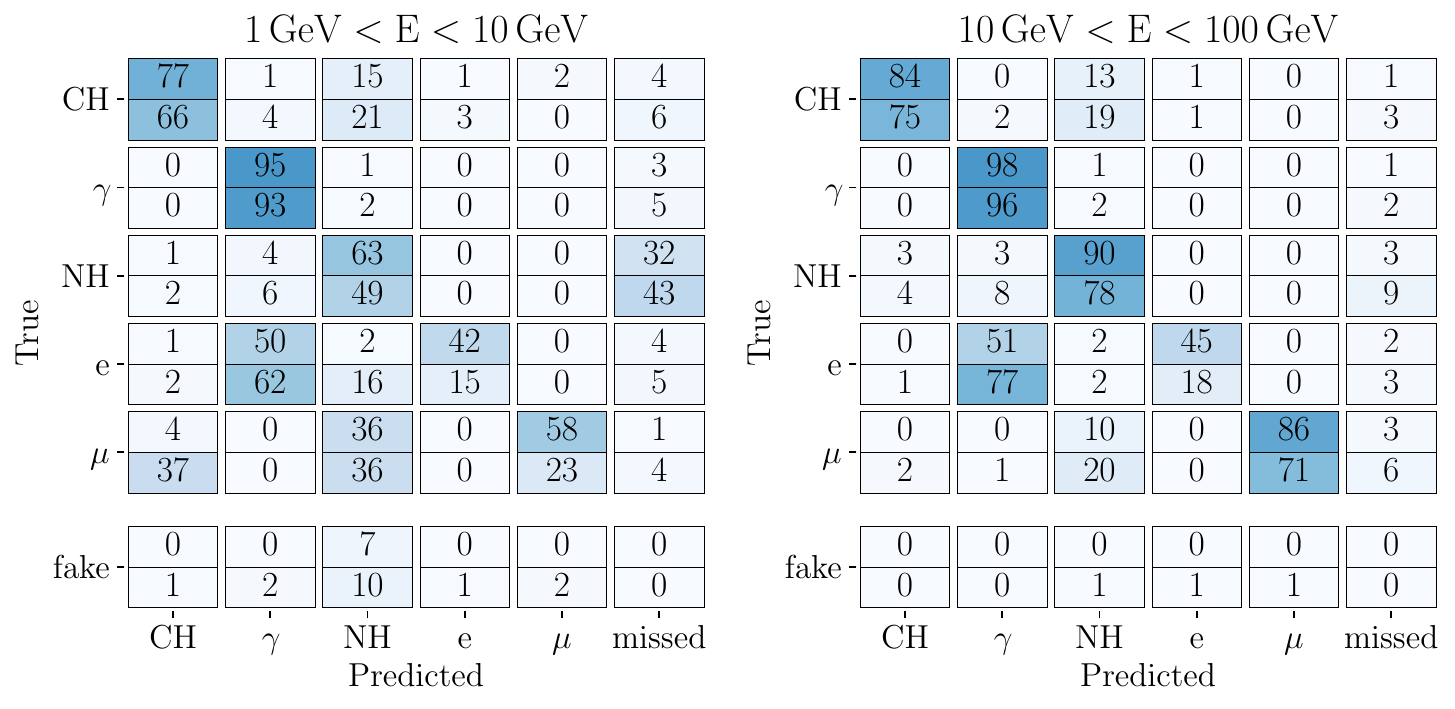}
    \caption{Confusion matrices comparing \hitpf (top entry of each tile (\%)) and \pandora (bottom entry of each tile (\%)) across two energy ranges: 1--10~GeV (left), and 10--100~GeV (right). Rows correspond to true particle type and columns to reconstructed type. Diagonal entries indicate correct identification, off-diagonal entries show mis-identification rates. The "missed" column represents the fraction of particles not reconstructed. The "fake" row indicates the percentage of reconstructed candidates in each class without a matched target particle. All values are normalized per row, except for the fake rates, which are normalized by column. 
    }
    \label{fig:confusion_matrix_hitpf_vs_pandora}
\end{figure*}

Figure~\ref{fig:confusion_matrix_hitpf_vs_pandora} shows confusion matrices comparing \hitpf and \pandora across the five particle classes in two energy ranges. The trends observed in the efficiency and fake rate distributions of Figure~\ref{fig:efficiency_fake_rate} are confirmed and can be discussed in more detail here.

For charged hadrons, even at high energies, a notable fraction is mis-reconstructed as neutral hadrons by both algorithms, more frequently by \pandora than \hitpf. This is primarily driven by inefficiencies in the tracking algorithm and represents an area of future improvement. For electrons, both algorithms show significant confusion with photons and
neutral hadrons, driven by secondary electrons with low tracking efficiency.
This misidentification occurs more frequently for \pandora than for \hitpf. For muons, both algorithms feature some degree of mis-reconstruction, although \hitpf achieves higher overall efficiency and much lower fake rate. For neutral hadrons, \hitpf achieves both higher reconstruction efficiency and lower confusion with photons than \pandora, while also reducing the fraction of missed particles. Overall, the diagonal is strongly populated for \hitpf, indicating efficient reconstruction. Across all energy ranges, \hitpf achieves higher efficiency and lower fake rate. 

\subsection{Energy and angular resolution}

\begin{figure*}[ht]
    \includegraphics[width=0.99\textwidth]{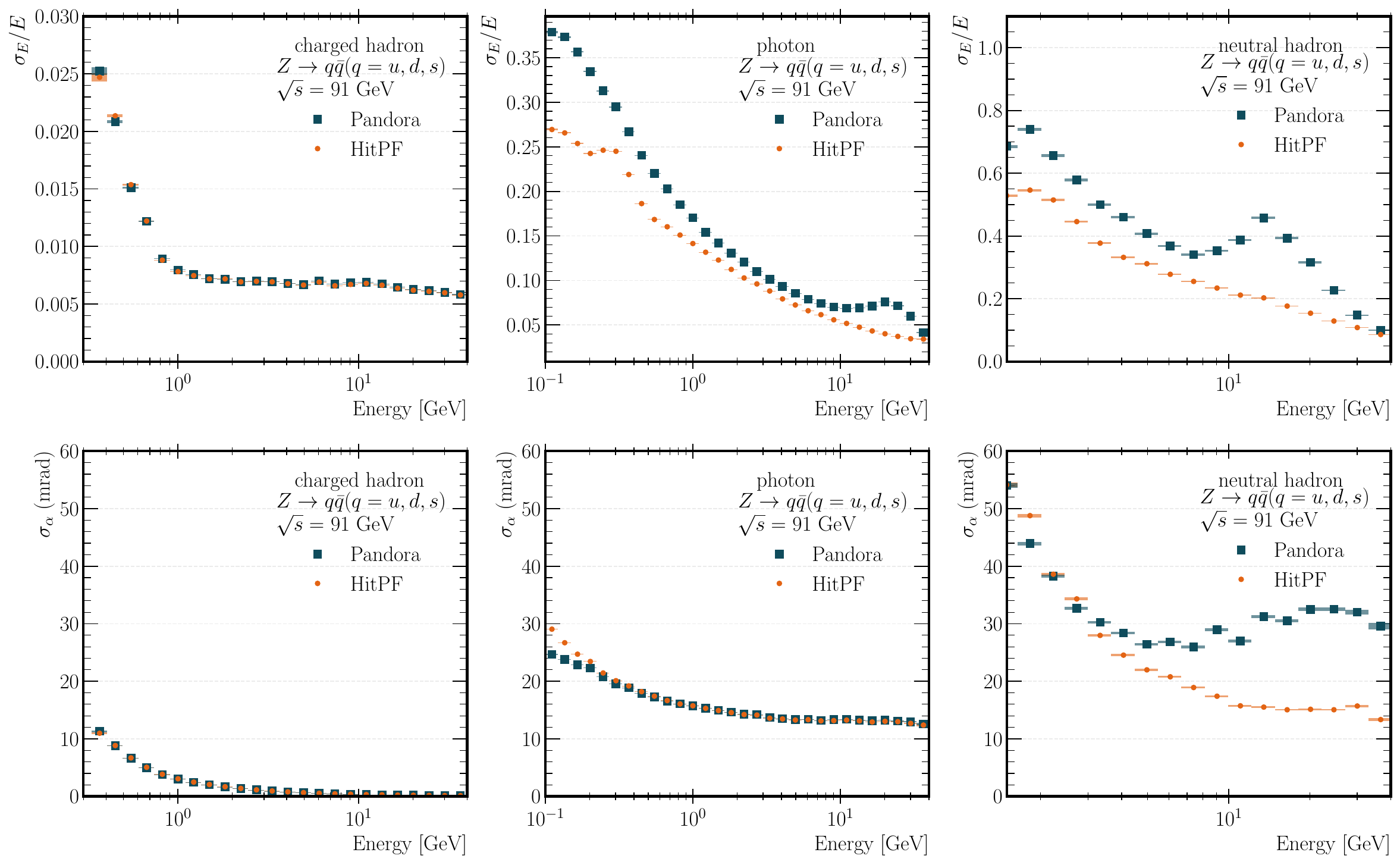}
    \caption{Energy resolution (top row) and angular resolution (bottom row) as a function of the true energy for charged hadrons (left), photons (centre), and neutral hadrons (right). The energy resolution is defined as the half-width of the 16th--84th percentile interval divided by the median of the $E_{\textrm{reco}}/E_{\textrm{true}}$ distribution. The angular resolution is defined as the 68th percentile of the $\alpha$ distribution, where $\alpha$ is the opening angle between the true and reconstructed particle directions.}
    \label{fig:fccee_energy_resolution}
\end{figure*}

Achieving the best possible resolution for each particle species is essential, since
energy and angular resolution of individual particles directly determine the invariant mass resolution of reconstructed composite objects, as will be shown in Section~\ref{subsec:eventlevel}. 

The energy resolution is computed as the half-width of the interval between the 16th and 84th percentiles divided by the median of the $E_{\textrm{reco}}/E_{\textrm{true}}$ distribution. This definition reduces to the standard deviation for Gaussian distributions but provides a robust metric for skewed or non-Gaussian tails. The angular resolution $\sigma_{\alpha}$ is defined as the 68th percentile of the $\alpha$ distribution, where $ \alpha$ is the opening angle between the true and reconstructed particle directions. The results are shown in Figure~\ref{fig:fccee_energy_resolution}.

For charged hadrons, \hitpf and \pandora achieve almost identical energy resolution by construction, since the particle momentum is obtained from standalone tracking in both cases. No calorimetric energy correction is applied to charged particles in the current implementation, as track momentum provides superior resolution below 50~GeV, the regime considered here. Combining track and calorimeter information for energy regression of charged hadrons is left for future work. The angular resolution is also comparable between the two algorithms, as it is dominated by the track direction measurement.

For photons and neutral hadrons, \hitpf achieves better resolution than \pandora, in particular for energies above 10~GeV, where \hitpf captures a larger fraction of the shower hits within each cluster. When neutral hadrons fall in the close vicinity of other particles, a frequent occurrence in the hadronic events considered here, \pandora tends to merge nearby showers into single clusters, resulting in overestimated energies. \hitpf disentangles nearby showers more effectively, reducing this effect.

\subsection{Event level observables}
\label{subsec:eventlevel}

Event level observables provide a comprehensive assessment of reconstruction performance, as they integrate the effects of single particle resolution, reconstruction efficiency, and fake rates into metrics directly relevant for physics analysis. Missed particles, fake contributions, and energy/direction mis-measurements all degrade the event invariant mass and visible energy resolution.

These observables directly determine the sensitivity of precision measurements of physical quantities. The visible mass spectrum is typically the observable used to extract a Higgs resonance peak over background. In the regime where the signal yield is small compared to background and the signal peak is approximately Gaussian, an improvement in mass resolution by a factor $f$ reduces the integrated luminosity required to achieve a given statistical precision by approximately the same factor.

\begin{figure*}[htbp]
    \centering
    \includegraphics[width=0.99\textwidth]{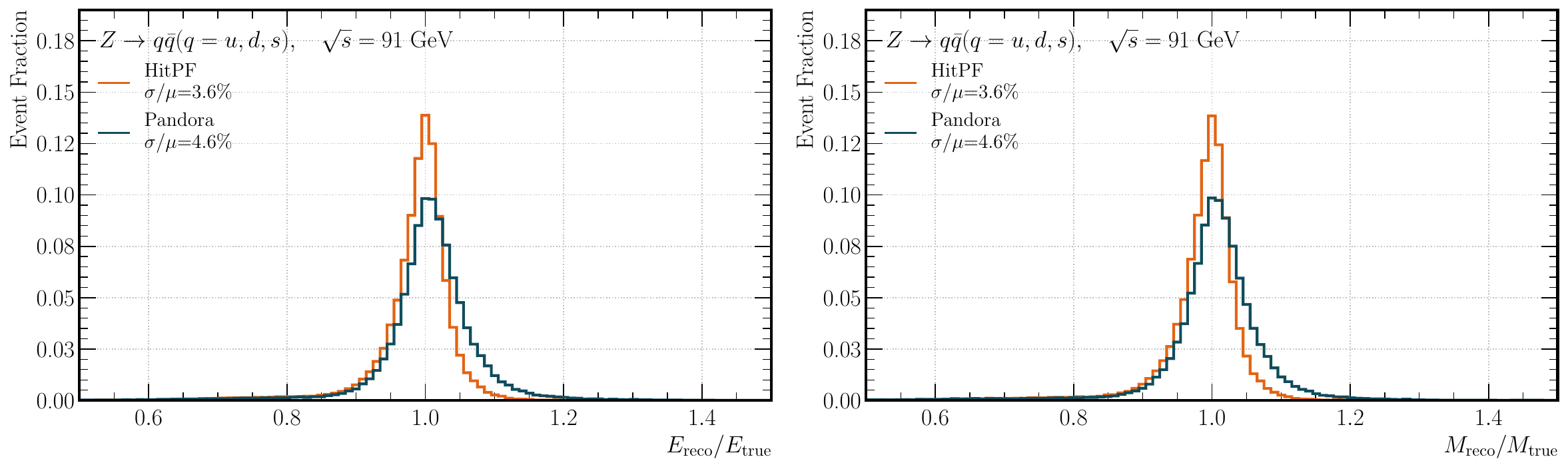}
    
\caption{Distribution of the reconstructed visible mass normalized to the true visible mass $M_{\textrm{reco}}/M_{\textrm{true}}$ (left) and reconstructed visible energy normalized to the true visible energy $E_{\textrm{reco}}/E_{\textrm{true}}$ (right) for \hitpf and \pandora. The median $\mu$ and relative resolution $\sigma/\mu$ are reported in the legend.} 
    \label{fig:fccee_mass}
\end{figure*}

Figure~\ref{fig:fccee_mass} shows the distributions of reconstructed visible energy and mass normalised to their true values. \hitpf achieves a 22\% relative improvement in both visible mass and energy resolution ($\sigma/\mu$) compared to \pandora.

\section{Discussion}
\label{sec:discussion}

We have presented \hitpf, an end-to-end particle flow reconstruction algorithm 
that maps calorimeter hits and charged particle tracks directly to particle objects 
without intermediate clustering stages or detector-specific tuning. The performance 
has been evaluated on simulated $Z \rightarrow q\bar{q}$ events in the FCC-ee CLD 
detector. \hitpf outperforms \pandora across all metrics: reconstruction efficiency 
in complex hadronic topologies improves by up to a factor of two at low momenta, 
neutral hadron efficiency increases by 10--20\% extending to higher energies, fake 
particle contributions are substantially reduced, and the visible mass resolution 
improves by 22\%.

Beyond the improvement in performance, a key advantage of this approach lies in its 
independence from detector-specific assumptions. Classical algorithms such as \pandora require extensive tuning of clustering parameters for each detector geometry, a process that must be repeated whenever the detector design changes, requiring significant expert effort. Instead, \hitpf learns reconstruction directly from simulation, enabling fast iterations during detector optimisation with minimal manual intervention. The input representation $(E, \vec{x}, s)$ can be employed across any tracking and calorimeter system, and adapting \hitpf to a new geometry requires retraining in approximately 48 hours on 4 NVIDIA H100 GPUs, a modest computational cost compared to the manual effort required to tune classical algorithms or, in the case of substantially different detector technologies, to design new reconstruction algorithms from scratch. The same framework can therefore be applied to other detector concepts under consideration for FCC-ee, enabling direct and consistent comparison of physics performance across designs.

The relative contributions of the geometric algebra representation, the object condensation training objective, and the density peak clustering to the overall performance improvement have not been isolated in the present study; systematic ablation is left for future work. Extension to hadron collider environments, where pile-up and higher multiplicities pose additional challenges, represents a natural next step. The code and models used to produce the results of this paper are made available at~\cite{garcia2026}.

\section{Acknowledgements}
We thank Brieuc François and the FCC software team for valuable comments and technical support, Josep Pata for early discussions on ML-based particle flow, Victor Breso for his comments on this manuscript, and Jan Kieseler for discussions on object condensation.
We acknowledge EuroHPC Joint Undertaking for awarding the project ID EHPC-AIF-2025SC02-040 access to MareNostrum5 at BSC, Spain.
This work was also supported by the Ministry of Education, Youth and Sports of the Czech Republic through the e-INFRA CZ (ID:EU2025D01-051).
We acknowledge the support of the ETH Zurich Institute for Particle Physics and Astrophysics for funding Gregor Krzmanc's stays at CERN.

\appendix

\section{Clustering Metrics}\label{app:clustering_metrics}

We present here the clustering performance prior to particle identification,
isolating the contribution of the hit-to-cluster assignment from the
subsequent classification step.

\begin{figure*}[htbp]
    \centering
    \includegraphics[width=0.99\textwidth]{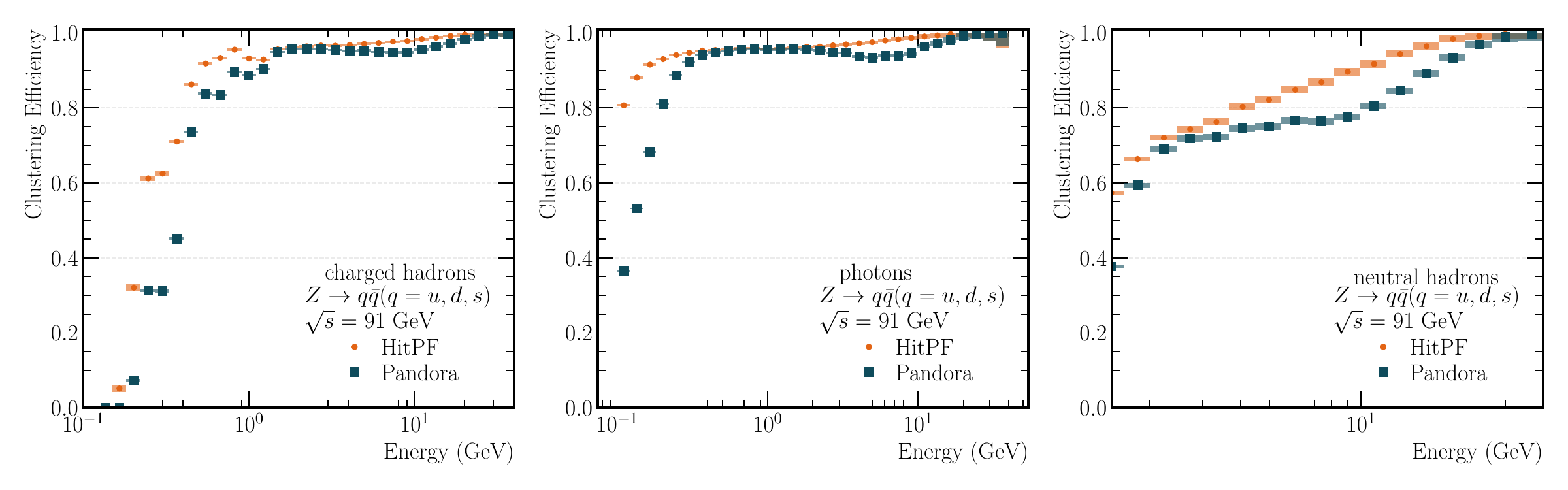}
    \caption{Clustering efficiency without particle identification for
    charged hadrons (left), photons (centre), and neutral hadrons (right) as
    a function of the matched MC particle energy.}
    \label{fig:clustering_metrics_appendix}
\end{figure*}

Figure~\ref{fig:clustering_metrics_appendix} shows the clustering efficiency
as a function of the matched MC particle energy for charged hadrons, photons,
and neutral hadrons. The clustering efficiency is defined as the fraction of
target particles matched to a reconstructed cluster, regardless of the
assigned particle type. \hitpf achieves higher clustering efficiency than
\pandora across all three particle species and over the full energy range,
with the largest gains observed for charged hadrons at low energies.

\begin{figure*}[htbp]
    \centering
    \includegraphics[width=0.99\textwidth]{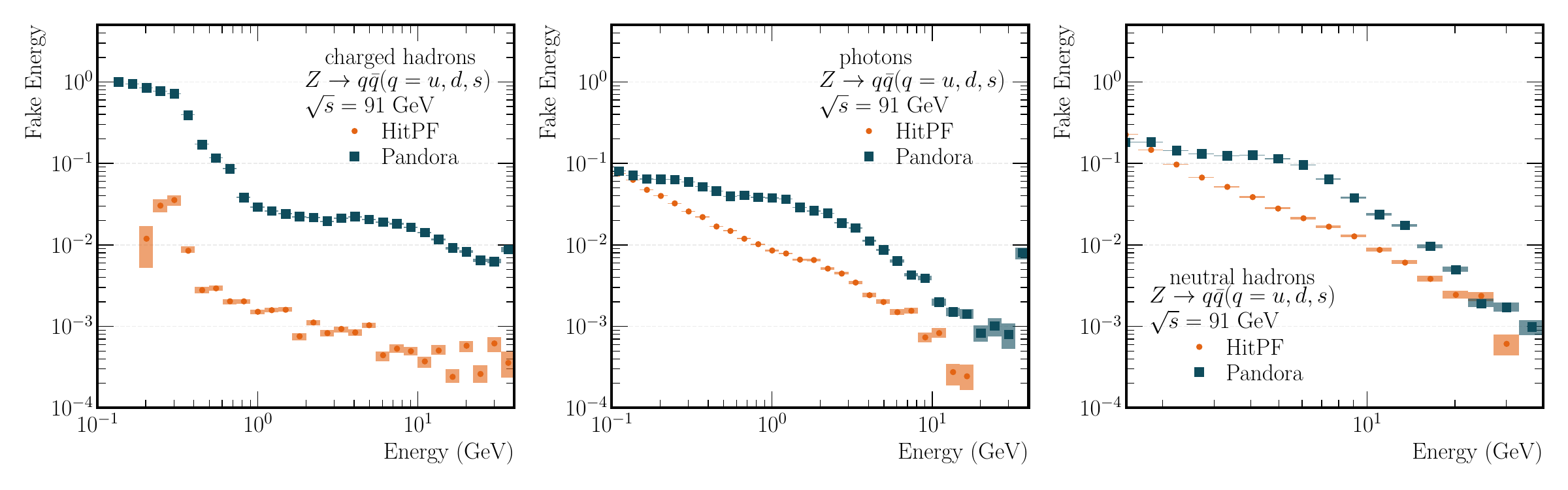}
    \caption{Fraction of reconstructed energy assigned to fake clusters for
    charged hadrons (left), photons (centre), and neutral hadrons (right) as
    a function of the reconstructed energy of the fake candidate. \hitpf is
    shown in blue and \pandora in orange.}
    \label{fig:fake_energy}
\end{figure*}

Figure~\ref{fig:fake_energy} shows the fraction of total reconstructed
energy assigned to fake clusters, clusters without a matched target
particle, as a function of the reconstructed energy. \hitpf produces
consistently lower fake energy than \pandora for all particle species,
confirming that the improvements observed in the particle identification
metrics of Section~\ref{sec:results} are already present at the clustering
level.

\section{Event Visualisation}\label{app:event_display}

We illustrate the qualitative differences between \hitpf and \pandora
through a full event display and a zoomed view of a particularly
instructive region.

\begin{figure*}[htbp]
    \centering
    \includegraphics[width=0.85\textwidth]{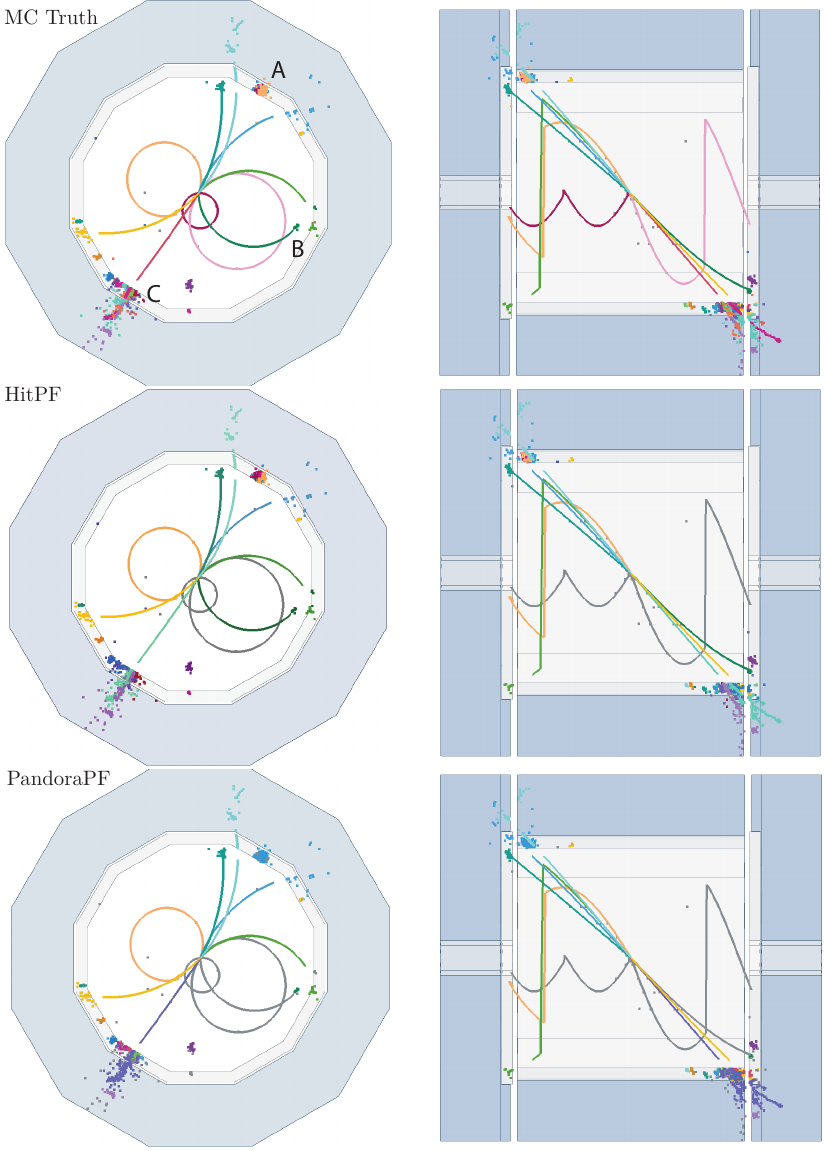}
    \caption{Visualisation of a $Z\rightarrow q\bar{q}$ event using \cite{dudarboh_mlpf_truth_ced_display_2026, holbe2012_ced_manual}. The left
    column shows the transverse view, the right column the longitudinal
    view. From top to bottom: MC truth hit assignment, \hitpf
    reconstruction, and \pandora reconstruction. Hit colours for \hitpf
    and \pandora correspond to the MC particle contributing the largest
    hit count to each reconstructed cluster. Three failure modes of
    \pandora are highlighted: A)~four nearby photons merged into a single
    35~GeV cluster (see Figure~\ref{fig:zoomA} for detail),
    B)~a charged particle whose track is not associated to the cluster,
    resulting in reconstruction as a neutral hadron with degraded energy
    resolution, and C)~a proton interaction in the tracker producing
    multiple secondary particles that enter the calorimeter, leading to
    double counting of energy.}
    \label{fig:visualization_event}
\end{figure*}

Figure~\ref{fig:visualization_event} shows the transverse and longitudinal
views of a representative $Z \rightarrow q\bar{q}$ event, comparing the MC
truth hit assignment with the reconstructions from \hitpf and \pandora. We
highlight three failure modes of the classical algorithm that are handled
more effectively by \hitpf.

\begin{figure*}[htbp]
    \centering
    \includegraphics[width=0.99\textwidth]{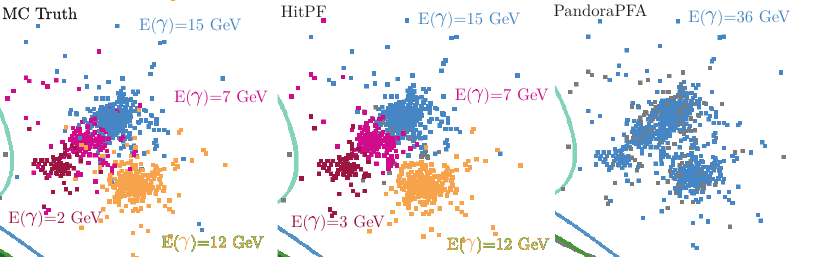}
    \caption{Zoomed view of region A from
    Figure~\ref{fig:visualization_event}, where four nearby photons are
    merged into a single cluster by \pandora. \hitpf correctly resolves
    the individual particles.}
    \label{fig:zoomA}
\end{figure*}

In region A, four nearby photons produce partially overlapping
electromagnetic showers. \hitpf correctly resolves all four particles, while
\pandora merges them into a single cluster with a reconstructed energy of
35~GeV, exceeding any individual photon energy. A zoomed view of this
region is shown in Figure~\ref{fig:zoomA}.

In region B, \pandora fails to associate a track with its corresponding
calorimeter cluster (shown in grey in the bottom row). As a result, the
charged particle is reconstructed as a neutral hadron, hence neglecting the superior
track momentum measurement and relying instead on the calorimetric energy
estimate with its poorer resolution.

In region C, a proton undergoes a nuclear interaction in the tracker
material, producing multiple secondary particles that enter the calorimeter.
\pandora associates one or more of the secondary showers with the original
track, while the remaining daughters contribute additional un-associated energy.
This results in partial double counting the original proton energy: the track momentum is used for the energy estimate, but the calorimetric deposits from the other daughters are
reconstructed as separate particles, inflating the total reconstructed
energy. \hitpf produces a more faithful representation of the event by
clustering the secondary showers independently of the original track.

\clearpage

\bibliographystyle{cms_unsrt}
\bibliography{references}

\end{document}